\documentclass{pnastwo}

\usepackage[dvips]{graphicx}
\usepackage{amssymb,amsfonts,amsmath}

\url{www.pnas.org/cgi/doi/10.1073/pnas.0709640104}
\copyrightyear{2009} \issuedate{Issue Date} \volume{Volume}
\issuenumber{Issue Number}

\begin{document}

\title{Confinement Effects on the Kinetics and Thermodynamics of
Protein Dimerization}

\author{Wei Wang$^\ast$, Wei-Xin Xu\affil{1}{National Laboratory of Solid
State Microstructure and Department of Physics, Nanjing University,
210093, China}, Yaakov Levy\affil{3}{Department of Structural
Biology, Weizmann Institute of Science, Rehovot 76100, Israel}, E.
Trizac\affil{4}{CNRS; Universit\'e Paris-Sud, UMR8626, LPTMS, Orsay
Cedex, F-91405, France}, \and P. G. Wolynes\affil{5}{Department of
Chemistry and Biochemistry, University of California at San Diego,
9500 Gilman Drive, La Jolla, CA 92093-0371, USA}}

\maketitle

\begin{article}
\begin{abstract}

In the cell, protein complexes form relying on specific interactions
between their monomers. Excluded volume effects due to molecular
crowding would lead to correlations between molecules even without
specific interactions. What is the interplay of these effects in the
crowded cellular environment? We study dimerization of a model
homodimer both when the mondimers are free or tethered to each
other. We consider a structured environment: Two monomers first
diffuse into a cavity of size  $L$ and then fold and bind within the
cavity. The folding and binding are simulated using molecular
dynamics based on a simplified topology based model. The {\it
confinement} in the cell is described by an effective molecular
concentration $C \sim L^{-3}$. A two-state coupled folding and
binding behavior is found. We show the maximal rate of dimerization
occurred at an effective molecular concentration $C^{op}\simeq 1m$M
which is a relevant cellular concentration. In contrast, for
tethered chains the rate keeps at a plateau when $C<C^{op}$ but then
decreases sharply when $C>C^{op}$. For both the free and tethered
cases, the simulated variation of the rate of dimerization and
thermodynamic stability with effective molecular concentration
agrees well with experimental observations. In addition, a
theoretical argument for the effects of confinement on dimerization
is also made.
\end{abstract}

\keywords{molecular crowding | dimerization | folding and binding |
effective molecular concentration | Arc homodimer monolayer}

\abbreviations{}

\dropcap {M}any biological functions depend on protein complexes or
multimeric proteins which must specifically form in a crowded
cellular environment. There are several types of protein complexes.
Homodimeric proteins consisting of two identical chains or monomers
with a symmetrical conformation are the most typical \cite{dimer}.
{\it In vitro} experiments show that the formation of dimeric
proteins, termed dimerization, may be described as two-state
\cite{Two-state} or three-state \cite{three-state}. Here, the term
two-state indicates that the folding and binding of monomers are
directly coupled, while the term three-state signifies that binding
starts from already folded monomers or that binding has a dimeric
intermediate. Since dimerization involves assembly of two monomers,
its rate should depend on the monomer concentration. That is, when
the separation distance of the monomers is large, the monomers
should diffuse close to each other first and then dimerize. For {\it
In vitro} experiments where there is only one kind of molecule
involved in general, the dimerization occurs easily when the
concentration is large.  {\it In vivo} dimerization of specific
monomers is more complicated than {\it in vitro} because cells are
rather crowded due to the presence of various
macromolecules\cite{crowding1,crowding2,crowding3,crowding4,hxzhou-minton2008}.
When the local concentration of the monomers is low, the monomers
take a long time to diffuse together, and the diffusion even may be
kinetically blocked by other molecules. This makes dimerization more
difficult. Nevertheless, when the local concentration of the
monomers is sufficiently high, dimerization occurs easily.

The concentration of total macromolecules in cytoplasm is estimated
to be $80\sim 200g/l$ \cite{crowding1,crowding2,crowding3} which is
approximately $1m$M (or $100\mu$M) if the averaged molecular weight
$\bar{m}=500\times 110$Da, i.e., 500 amino acids (or 5000 amino
acids) in average for a macromolecules, is assumed. Obviously,
crowding must lead to excluded volume effects
\cite{crowding4,hxzhou-minton2008,excluded1,excluded2,excluded3,excluded4,excluded5}
which can be described using an effective concentration of the
reacting molecules. Crowding can preferentially destabilize the
balance between reactants and products, and makes the association
reactions highly favored. It has been suggested that association
constants under crowded conditions could be several orders of
magnitude larger than those in dilute
solutions\cite{crowding2,crowding3,crowding4,hxzhou-minton2008}. At
the same time, crowding causes a decrease in the diffusion rate of
molecules by a factor in the range of $3\sim 10$
\cite{crowding2,crowding3,verkman}.

The translational diffusion of molecules in the cell is a kinetic
process which can be described using Brownian dynamics
\cite{verkman,northrupjcp1984}. Dimerization, ultimately, involves
the intimate contact between two specific monomers, a local dynamic
process. Previously, the simultaneous folding and binding of a
number of homodimers has been theoretically studied using topology
based models (Go-like) by adding a covalent linkage between the two
monomers of the dimer \cite{dimer-go1,dimer-go2}. Such studies may
be directly related to the {\it in vitro} situation. To study the
{\it in vivo} dimerization of homodimeric proteins, both
interactions between the monomers and those between the monomers and
other macromolecules must be considered. Including crowding effects
in such studies may provide some useful insights into the formation
of various protein complexes and protein-protein interactions, and
thus enable one to understand intracellular protein networks, and to
design protein complexes that could act as pharmacological
inhibitors.

Here, we study the dimerization of two monomers encapsulated in a
cavity with size $L$ which mimics the crowding in cell by an
effective molecular concentration $C \sim L^{-3}$. We study both the
thermodynamics and kinetics. The diffusion of the monomers into the
confined space is described by a Brownian dynamics. Dimerization
depends on the size $L$, or the effective concentration $C$. There
the model predicts a maximal rate of dimerization at an optimal
confined space size $L^{op}=22$ (with unit $3.8$\AA). Such an
optimal $L^{op}$ corresponds to an effective concentration
$C^{op}\sim 1m$M which is of the order of the macromolecular
concentration in cells \cite{crowding1,crowding2,crowding3}. This
suggests a possibility that the rate of dimerization and the
concentration of various macromolecules in cells may have been
optimized by evolution. Based on the changes of the conformational
and translation entropies due to the confinement, we show that there
is a scaling behavior for the heights of free energy barriers for
binding and the folding transition temperatures with the cavity
sizes.

\section{Results and Discussions}

\subsection{Molecular Crowding and Molecular Diffusion}
Suppose that in a cubic box with size $L_b=1000\AA$, corresponding
to a small compartment of the cell, there are about 1000 molecules.
Among these only two specific monomers can form a dimer. The
effective molecular concentration is $C_{e}=100g/l$ given an average
molecular weight $\bar{m}=55$KDa (i.e., $\sim 500$ amino acids).
That is, $1000=C_{e}\times L_{b}^{3}$. All molecules diffuse
randomly in the large box. Once the distance between two monomers
reaches a smaller value, $L$, i.e., the monomers diffuse into a
small confined space (Fig.1{\it A}), there they can form a dimer by
folding and binding. The diffusion time is simulated using a
Brownian dynamics (see the Methods), and a monotonic decrease of the
time versus the size $L$ is obtained (Fig.1{\it B}).

In the Brownian dynamics each monomer is modeled as a single
particle mimicking the protein chains. Clearly, the diffusion of
such particle could not fully describe the case of the protein
chains since the protein chains are soft and may change their
conformations during the diffusion. However, for the sake of
simplicity, an effective friction for the particles could be used to
model the diffusion of the protein chains. Previously, a friction
coefficient $\gamma_{a} \sim 0.05$ was used for a single amino acid
with size $a$ \cite{Thirumalai1996}. For the subunit of Arc
repressor (with 53 amino acids) studied in this work, a
corresponding friction coefficient is estimated to be $\gamma \sim
0.19$ since the size of the Arc monomer is approximately
$53^{1/3}a$. Here, a spherical conformation for the monomer is
assumed. Note that for a protein with $500$ amino acids a friction
coefficient is $\gamma \sim 0.4$ [If a friction coefficient
$\gamma_{a}=0.2$, which was argued to be a factor of 10 larger than
the measured value for amino acids in water\cite{pnas-Nymeyer}, is
set, one has $\gamma \sim 1.6$]. Thus a value of $\gamma \sim 0.1$
is in the reasonable range to model the kinetics for protein Arc. To
show the effect of friction coefficient on diffusion rate of the
monomers, several cases with different $\gamma$ values are simulated
(see Fig.1{\it B}). Clearly, the diffusion slows down when the
friction is large. It is noted that such simplified diffusion of
Brownian particles is approximate but reasonable when the sizes of
the conformations of the two protein chains can be negligible when
compared with the inter-chain distance between them. Obviously, if
the density of the specific monomers is high, the diffusion time
will be short and the dimerization events will be more frequent. In
the present work we only put two such monomers to model the least
case. It is also worth noting that if the size of confined space
$L$ is roughly $5\sim 6$ times the radius of gyration $R_g$ of the
monomers, the monomers need to diffuse further within the confined
space. Here a value $5\sim 6$ is set since unfolded monomeric chain
is extended \cite{excluded3}.

\begin{figure}
\centerline{\includegraphics[width=.4\textwidth]{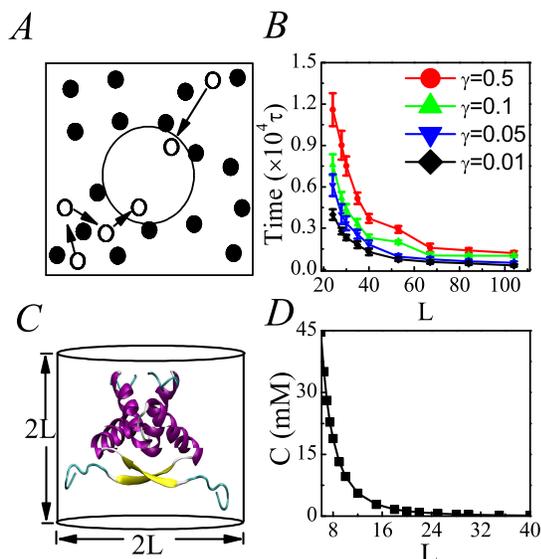}}
\caption{ Molecular crowding and confinement. ({\it A}) A schematic
particle model for the molecular crowding. Two monomers are
distributed randomly in a box and diffuse into a confined space with
size $L$. ({\it B}) The time of any two monomers diffusing closely
with a separating distance $<L$ versus the size $L$ for four
different friction coefficients. ({\it C}) The Arc dimer (PDB
structure 1arr) confined in a cylindrical cavity. ({\it D}) The
cavity size $L$ versus the effective molecular concentration $C\sim
1/(2\pi L^{3})$} \label{fig1}
\end{figure}

\subsection{Model Protein and Confinement} The homodimer
studied in this work is the Arc repressor of bacteriophage P22 which
consists of two chains each containing 53 residues and has a
symmetrical native structure \cite{BREG}. Clearly, Arc is taken as
the model protein because it is small and has been studied
experimentally. In a crowded circumstance, the confined space is
taken as a cylindrical cavity (Fig.1{\it C}) and its diameter $2L$
and height $h$ are set equal to each other. The size of the cylinder
is related to the effective molecular concentration by $C\sim
1/(2\pi L^3)$ if every cylinder only contains two molecules.
Obviously, a big cylinder or weak confinement corresponds to a low
$C$ and {\it vice versa} (Fig.1{\it D}). Thus, two Arc monomers may
fold and bind in such a cavity, modelling the dimerization of a
homodimer within a crowded cell. Such behaviors are simulated based
on the molecular dynamics using the Go-like potentials (see the
Methods).

 The confinement is modeled by a cylinder which was previously
used to study the confinement/crowding effect on protein folding
\cite{excluded1,excluded4,excluded5,Ziv05pnas}. Simulations using
spherical space provide that using different confinement shapes will
not qualitatively change the results
\cite{Klimov02pnas,excluded2,Mittal2008}. Furthermore, a study on
the molecular crowding effect on folding of globular proteins
suggested that to depict a rather crowded {\it in vivo} environment,
the optimal cavity to host protein molecule may be cylindrical
\cite{excluded3}.

\subsection{Two-state Behavior} Some features of the dimerization
trajectories of the Arc dimer confined in a cavity with $L=20$ (or
$C\sim 1.2m$M) at the related transition temperature $T^{L}_f$ are
shown in Fig.2. The typical time evolution of the native contacts of
chain-A and chain-B ($Q_A$ and $Q_B$) and the interfacial native
contacts ($Q_{AB}$) and the separating distance $d$ between the
centers of mass of two chains is shown (Fig.2{\it A-B}). One can see
clearly that the folded state (with $Q_A$ or $Q_B \sim 0.9$) emerges
only when the interface is formed (with $Q_{AB}\ge 0.85$).
Interestingly, $d$ varies between the folded and unfolded states.
The free energies of the folding and binding process projected onto
three different sets of reaction coordinates show the most populated
states, i.e., folded chains with a well formed interface (both $Q_A$
and $Q_B\sim 0.9$, and $Q_{AB}\sim 0.85$) and the unfolded chains
without binding (both $Q_A$ and $Q_B\sim 0.5$, and $Q_{AB}< 0.1$)
(Fig.2{\it C-D}). Note that the interfacial native contacts
$N_{AB}=143$ are almost twice as numerous as the intra-chain native
contacts $N_{A}$ (or $N_{B}$) $=77$, thus energetically, the states
with $Q_A$ and $Q_B\sim 0.5$ can still be referred as the unfolded
states. These indicate that the folding and binding occur in a
cooperative two-state manner, consistent with previous {\it in
vitro} experimental observation
\cite{Two-state,experiment5,experiment6}, and also with earlier
simulations for linked chains \cite{dimer-go1,dimer-go2}.

\begin{figure}
\centerline{\includegraphics[width=.4\textwidth]{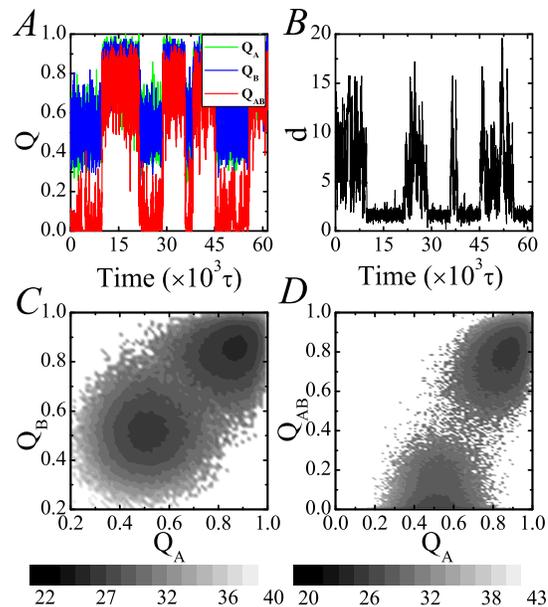}}
\caption{The features of dimerization at $T^{L}_f$ within a cavity
of $L=20$. ({\it A}) The time evolution of the native contacts:
$Q_A$ for monomeric chain-A in green (or $Q_B$ for monomeric chain-B
in blue), and $Q_{AB}$ for the interface in red. ({\it B}) The time
evolution of the separated distance $d$ between the centers of mass
of two monomers. ({\it C}) The free energies projected onto $Q_A$
versus $Q_B$. ({\it D}) The free energies projected onto $Q_A$
versus $Q_{AB}$. } \label{fig2}
\end{figure}

\begin{figure}
\centerline{\includegraphics[width=.4\textwidth]{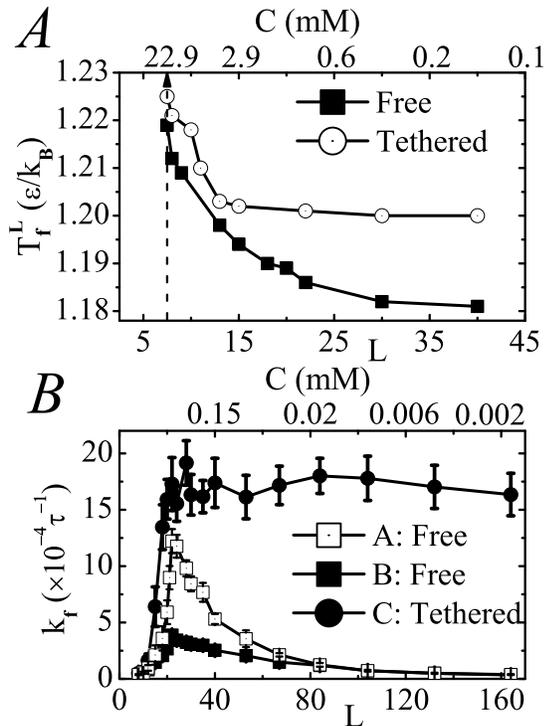}}
\caption{The features of dimerization. ({\it A}) The folding
transition temperature $T^{L}_f$ versus the cavity size $L$ for two
free and tethered monomers. ({\it B}) The dimerization rate averaged
over $100$ trajectories at $0.85T^{L}_f\sim 1.0$ versus the cavity
size $L$.  Curve-A shows the case without the global diffusion for
two free monomers, and curve-B shows the case with the global
diffusion for two free monomers at $\gamma =0.1$. Curve-C shows the
case for two tethered monomers. The related concentrations are
listed in the upper x-axes in both Fig.3{\it A-B}.} \label{fig3}
\end{figure}

\subsection{Effects of Concentration on Stability} To
study the influence of the various effective molecular
concentrations $C$ on the dimerization, the transition temperatures
$T^{L}_f$, which characterizes the thermodynamic stability of the
dimer (high value of $T^{L}_f$ means high stability) are obtained.
In Fig.3{\it A}, it is shown that the value of $T^{L}_f$ decreases
monotonically as $L$ increases (or $C$ decreases), implying that
small $C$ or large space results in low value of $T^{L}_f$ or low
thermodynamic stability. Experimentally, both urea and thermal
denaturation showed that the stability of the Arc dimer is low at
low protein concentrations \cite{Two-state,experiment5,experiment6}.
Experiments on other dimeric proteins have also showed that high
concentration improves the thermodynamic stability
\cite{experiment3,experiment4}. Our results are clearly consistent
with these experimental observations. The value of $T^{L}_f$ at
$C=22.9m$M (or $L=7$) increases about $4\%$ with respect to that of
the case of confinement-free, i.e., $T_f^{bulk}=1.18$ defined
roughly at $C=1\mu $M. Obviously, such a big enhancement in the
thermodynamic stability is due to the crowding effect or confinement
which reduces the conformational and translational entropies of the
unfolded states of the two monomers more than it affects the native
dimer thus making the unfolded states unstable (see an argument in
the final part). Note that the dimerization cannot occur if the
confined space $L<7$ (see Fig.3{\it A}). This relates to a too
crowding case for the monomers to perform their folding and binding.

\subsection{Effects of Concentration on kinetics} The
effect of concentration on the kinetics of dimerization is also
reflected in the rate of dimerization by incorporating the
diffusion, folding and binding processes together (Fig.3{\it B}).
The rate $k_f$ changes nonmonotonically as $L$ increases (or $C$
decreases), showing an optimal maximum at $C^{op} \sim 1m$M which is
relevant to the macromolecular concentration in cells (see curve-A
and curve-B in Fig.3{\it B}). Here the rate $k_f$ is in inverse
proportion to the summation of the time for the two monomers to
diffuse into the confined space and time for assembly of two
monomers within the confined space. Note that the assembly of two
monomers within a confined space may include the local diffusion if
the initial distance between the two monomers is large. Clearly,
here the diffusion of two monomers in the cavity is simulated by the
motion of two poloymeric chains (Fig.1{\it C}) not of two point
particles (Fig.1{\it A}).

In Fig.3{\it B}, three cases, namely those for the dimerization of
two monomers with and without diffusion and for the single tethered
mutant, are shown. Curve-A shows the case without nonlocal
diffusion, which describes a situation of high local concentration
of the monomers. It is found that when $L$ is small (or $C$ is
high), the dimerization is slow and quite difficult since the
conformational space for the chains to search is limited. As $L$
increases, the dimerization becomes easier and faster. However, when
$L$ is too large, the conformational space becomes very big and the
chains now must spend much time in finding the folded state,
resulting in slow dimerization. Thus, there exists an optimal size
for the confinement, or an optimal effective concentration $C^{op}$.
For $C<C^{op}$, the rate $k_f$ monotonically decreases as $C$
decreases (Fig.3{\it B}). When $C$ is low enough, the rate $k_f$
depends linearly on $C$ in agreement with the experimental
observation\cite{experiment5}. As shown by curve-B, a similar change
of dimerization rate is also observed when the nonlocal diffusion is
taken into account. Since the diffusion time decreases monotonically
as the size of confined space $L$ increases (Fig.1{\it B}), the
decrease of dimerization rate becomes slower. However, there still
exists an optimal size of the confinement, or an optimal effective
concentration having about the same value of $C^{op}$ obtained for
the case without the nonlocal diffusion. The physical origin for
such a behavior is basically the same as the case without diffusion,
and the nonlocal diffusion only increases the total time of the
dimerization when two monomers are further separated. Actually,
curve-B is related to rather rigorous environment since the local
concentration of the monomers is low (only two monomers among 1000
molecules are assumed) and the averaged separation distance is
large. Clearly, if the local concentration of the monomers is not so
low or the monomers are co-located, the effects of global diffusion
are smaller. As a result, a curve of dimerization rate should be
bounded by the curves - A and -B.

\subsection{Effect of Confinement for Tethered Mutant}
Clearly, for the tethered mutant, i.e., when the two chains of the
Arc dimer are linked together, the thermodynamic stability is higher
than for the non-tethered case, especially when $L\ge 15$ (see
Fig.3{\it A}), and the rate of dimerization shows a plateau when
$C<C^{op}$ (Fig.3{\it B}). Again this agrees with experimental
observations that tethering two subunits of a dimeric protein
significantly enhances both the thermal stability of the dimer and
the rate of dimerization \cite{experiment6,experiment7}. The
physical reason is that the tethering reduces significantly the
conformational and translational entropies of the two tethered
chains, resulting in the reduction of the search time in the
unfolded ensemble and destabilization of the unfolded states. In
addition, the two chains of Arc dimer would not need to diffuse much
to be close to each other because they are already linked together.
Therefore, it takes them less time to complete the folding and
binding in comparison to the non-tethered case especially for large
confined spaces. Obviously, more time is needed for diffusion as the
available space of two monomers grows larger. It is worth noting
that the tethered case actually is related to a rather crowded case
of non-tethered monomers, and gives an effective concentration
$C_{e}=2.7m$M for the non-tethered Arc dimers
here\cite{experiment6}. This is relevant to the optimized effective
concentration $C^{op}\sim 1m$M.

\subsection{Free Energy Profiles of Folding and Binding}
To characterize the folding and binding of the two chains, we
calculate the free energy profiles for both processes, respectively.
As shown in Fig.4{\it A}, for a case of $L=20$ (or $C\sim 1.2m$M),
the height of free energy barrier $\Delta G_b^{\ddag}$ for binding
is about $4.7\epsilon$ which is much larger than that for folding of
the monomeric chains, i.e., $\Delta G_f^{\ddag}\sim 1.7\epsilon$.
This suggests that the binding is a dominant rate-limiting step in
the dimerization. Interestingly, it is found that the value of
$\Delta G_b^{\ddag}$ increases when $L< 22$ (or $C>1m$M) and then is
saturated to $5.0$ when $L\ge 22$ (or $C\le 1m$M) (Fig.4{\it B}).
However, the variation of $\Delta G_f^{\ddag}$ for the monomeric
chain is rather small (Fig.4{\it C}). Therefore, the crowding effect
mainly influences the binding rather than the folding of the
monomers. The rate limiting step, the binding of two monomers, also
requires overcoming frustrated polar interactions or non-native
contacts formed at the interface between two monomers in a
relatively hydrophobic environment \cite{Waldburger1996pnas}.

To further understand the dimerization of the two chains, the free
energy profile as a function of $\Delta d$, the distance between the
centers of mass of two chains shifted by subtracting the native
separation distance, is shown in Fig.4{\it D}. Free energy funnels
can be clearly seen for three cases. As an example for the case of
$L=14$, a deep well around $\Delta d \sim 0$ corresponds to a quite
stable binding or a localized state of the two chains. Note that the
effective attraction is short ranged, and is similar to that of the
binding between the ligand and receptor. It is also seen that the
two chains have weak or even no interaction in a certain range of
$\Delta d\sim 12$ . However, due to the repulsive interaction
between the protein chains and the cavity wall representing the
excluded volume effects of other protein, the free energy increases
when $\Delta d=\Delta d^{*}>16$. Thus we see the dimerization is
cooperatively guided by the binding and confinement quite naturally.
For the various cavity sizes, the ranges with weak interactions and
the values of $\Delta d^{*}$ are different, indicating that the
slopes of the free energy profiles are different. The presence of a
free energy funnel allows the dimerization to be stabilized by
confinement. This is very similar to that of a free energy of
protein-ligand binding obtained theoretically and also to the force
measured for the ligand-receptor association and dissociation
\cite{Woopnas2005,Moy}.

The dimerization reaction under confinement conditions investigated
in this study by the native topology-based model
\cite{Shoji-pnas,dimer-go1,dimer-go2} focuses on the effect of the
confinement on the configurational and translational
entropy\cite{excluded2}. In the cell, confinement and crowding
encapsulate the monomers closely and facilitate binding. It is
possible that the cavity has another role besides restricting the
available volume of protein motions and dynamics. For example, other
effects can arise due to interactions of the protein with the walls
of the cavity or due to intra- or inter-monomeric non-native
interactions.

\begin{figure}
\centerline{\includegraphics[width=.4\textwidth]{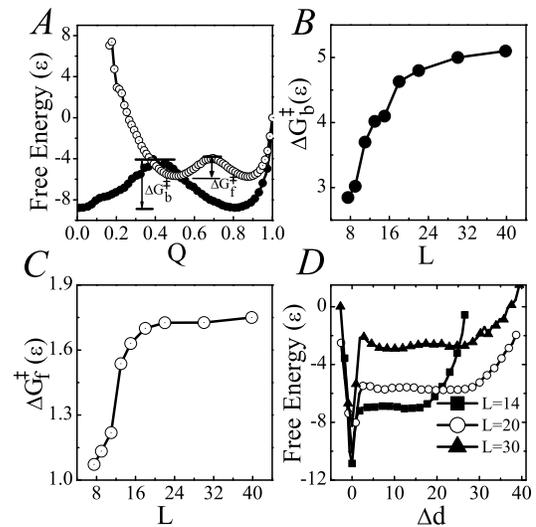}}
\caption{ The free energy profiles and barriers for folding and
binding. ({\it A}) The free energy for the interface as a function
of $Q_{AB}$ (solid circles), and the free energy for the monomeric
chain as a function of $Q_A$ (open circles). ({\it B}) The height of
free-energy barrier for the dimer $\Delta G^{\ddagger}_b$ (marked in
Fig.4{\it A}) versus $L$. ({\it C}) The height of free-energy
barrier for a monomeric chain $\Delta G^{\ddagger}_f$ (marked in
Fig.4{\it A}) versus $L$. ({\it D}) The binding free energy between
two monomers as a function of $\Delta d$.}\label{fig4}
\end{figure}

\begin{figure}[b]
\centerline{\includegraphics[width=.4\textwidth]{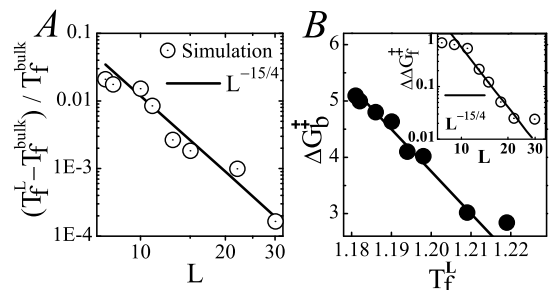}}
\caption{ Scaling behavior of folding and binding. ({\it A}) The
scaling of $\ln (T^{L}_{f}-T_{f}^{bulk})/T_{f}^{bulk}$ with $L$ for
two tethered monomers. The data are taken from Fig.3{\it A} (open
circles) and the line represents the theoretical argument
$(T^{L}_{f}-T_{f}^{bulk})/T_{f}^{bulk} \sim L^{-15/4}$. ({\it B})
The barrier height $\Delta G^\ddagger_b$ versus the folding
temperature $T_{f}^{L}$ of two free monomers confined in space with
$L$. The data are taken from Fig.4{\it B} (solid circles). The line
in the main graph is a guide to the eye. Inset: The changes of
barrier heights of the single monomer $\Delta\Delta G^{\ddagger}_f$
scaled with the size $L$. The open circles show the data from
Fig.4{\it C}, and the line shows the scaling $\Delta\Delta
G^{\ddagger}_f \sim L^{-15/4}$.} \label{fig5}
\end{figure}

\subsection{Theoretical Interpretation of Confinement} It
is well known that the folded state corresponds to a compact
conformation, while the ensemble of unfolded states has a huge
number of extended conformations. Thus, confinement primarily
affects the free energy of the unfolded states through the
conformational entropy. This effect can be quantified based on the
theory of polymers with excluded volume \cite{deGennes}. From the
scaling arguments, the conformational entropy cost reads
\cite{deGennes,Luijten,Raphael,excluded2} $\Delta S^{c}_{u}/k_B \,
\propto \, -N^{9/4} (a/L)^{15/4}$ where $u$ denotes the unfolded
states, $S^{c}$ the conformational entropy, $N$ the number of
residues (or beads) with size $a$ of the beads in a chain, and $L$
the size of the confined space. The exponent $15/4$ is more
generally equal to $3/(3\nu -1)$ where $\nu=3/5$ is the Flory
exponent. In addition, since at the folding transition temperature
the free energy differences between the unfolded states and native
state $\Delta G=G^{u}-G^{n}$ for both cases with and without
confinement are zero, we have a relationship between the folding
temperatures and the entropies as $ T^{bulk}_{f}
(S^{bulk}_{u}-S^{bulk}_{n}) = T^{L}_{f} (S^L_{u}- S^L_{n})$ where
the superscript $L$ and $bulk$ indicate cases with and without the
confinement. Thus, we have $(T^L_{f}-T^{bulk}_{f})
(S^{bulk}_{u}-S^{bulk}_{n}) = T^{L}_{f}[(S^L_{n}-S^{bulk}_{n})-
(S^L_{u}-S^{bulk}_{u})]= T^{L}_{f}(\Delta S_{n} - \Delta S_{u})$. In
general, we have both the conformational and translational parts for
$\Delta S$, i.e., $\Delta S = \Delta S^{c} + \Delta S^{t}$.

For the tethered case, two monomers actually become a ``long''
single chain, and their contributions of translational entropies to
$\Delta S$ are cancelled in a first approximation. Only their
contributions to conformational entropies remain. Thus we have
$(T^L_{f}-T^{bulk}_{f})/T^L_{f} \propto - \Delta S_{u}^{c}/k_B
\propto L^{-15/4}$. Since the relative shift
$(T^L_{f}-T^{bulk}_{f})$ is quite a bit smaller than unity, we have
$(T^L_{f}-T^{bulk}_{f})/T^{bulk}_{f} \propto L^{-15/4}$. As plotted
in Fig.5{\it A} for simulation data of Fig.3{\it A}, a well
agreement can be seen. For the case of two free monomers, the
process of folding and binding, i.e., a process $1+2 \to 12$,
involves the loss of one chain or monomer, and the translational
entropies correspondingly cannot be cancelled. For the
unfolded/unbound states with 2 chains, this contribution changes
with the confined volume $V\simeq L^3$ as $2 \log V$, whereas it is
only $\log V$ for the folded/bound state. Thus, we have
$(T^L_{f}-T^{bulk}_{f})/T^{bulk}_{f} \propto -\log V - \Delta
S_{u}^{c}/k_B$. The logarithmic term explains why in the $T^L_{f}$
versus $L$ plot (Fig.3{\it A}), the curve for two monomers does not
seem to converge towards a plateau while that for the tethered case,
where only the algebraic term is present, the curve does saturate at
large value of $L$.

It is clear that the transition state between the unfolded and
folded states is an ensemble with a non vanishing conformational
entropy due to its smaller spatial extension than the ensemble of
unfolded states. The transition state ensemble is less sensitive to
confinement, so its conformational entropy is not affected very much
by confinement. When the system is confined at a given temperature,
say at $T=T^{bulk}_f$, the relative positions of free energies of
the folded and transition states are not affected, while the
unfolded state is destabilized. As a result, one then needs to
increase the temperature by an amount $T^{L}_{f}-T^{bulk}_{f}$ to
reach the folding temperature. The transition state is stabilized by
an amount proportional to $T^{L}_{f}-T^{bulk}_{f}$. Thus, the
barrier for folding $\Delta G_b^{\ddag}$ should decrease linearly
with $T^{L}_{f}$. Such an expectation is consistent with our
simulation data as shown in the main plot in Fig.5{\it B} where a
linear behavior is observed. The inset which shows the difference
between the bulk barrier and that at a given confinement is thus an
indirect way to check the above linear relation between barrier for
folding and folding temperature shift.

\section{Conclusion}

A model of  confinement effects on dimerization of a typical
homodimeric protein was studied. It was found that both the
thermodynamics and kinetics of the dimerization are affected
significantly by the effective molecular concentration characterized
by the size of cavity. The thermodynamic stability of the dimer can
be enhanced and the dimerization can be accelerated as the
concentration $C$ increases. An optimal value of $C^{op}\simeq
1.0m$M is obtained. This value is of the order of the concentration
of macromolecules actually found in cells. The confinement and
binding enhance the folding funnel, stabilizing the dimerization of
two monomers.

\begin{materials}
\section{Methods}

\subsection{Molecular diffusion}
The diffusion of the molecules (i.e., particles) in a box is
simulated using a Brownian dynamics as $m^{p}_{i}{\bf \dot
v_{i}}(t)={\bf F_{i}}(t)-\gamma {\bf v_{i}}(t) +{\bf \Gamma}_{i}(t)$
\cite{Thirumalai1996}. Here, ${\bf v}$, ${\bf \dot v}$ and $m^{p}$
are the velocity, acceleration and mass of the particles,
respectively. The subscript index $i$ runs from $i=1$ to $i=2$ for
two specific particles or two monomers of the dimer, and from $i=3$
to $i=1000$ for all other particles in the box. For the sake of
simplicity, all particles are taken to be identical. That is, all
the sizes are equal to approximately as $53^{1/3} a$ and the masses
are $m^{p}=53m$ since the Arc monomer has 53 amino acids. Here the
size and mass of an amino acid are $a$ and $m$, respectively. ${\bf
F_{i}}$ is the force arising from the interaction between the
particles. A hard-core repulsive interaction between the particles
and also between the monomer and particles is set as
$V(r)=(\sigma^{p}_{0}/r)^{12}$ where the hard-core radius of
particle is $\sigma^{p}_{0}= (53)^{1/3}4.0$\AA, and $r$ is the
distance between the particles. And an attractive interaction with
the 12-10 Lennard-Jones (LJ) potential between the two monomers is
set as $V(r)=5(\sigma^{p}_{0}/r )^{12} - 6(\sigma^{p}_{0}/r)^{10}$.
 {\bf $\Gamma$} is
the white and Gaussian random force modeling the solvent collision
with the standard variance related to temperature by $\langle{\bf
\Gamma }(t){\bf \Gamma }(t')\rangle=6\gamma k_B T\delta (t-t')$
where $k_B$ is the Boltzmann constant, $T$ is absolute temperature,
$t$ is time, and $\delta (t-t')$ is the Dirac delta function. Four
values of the friction coefficient from $\gamma =0.01$ to $0.5$ are
used in our simulations (see Fig.1{\it B}). The temperature is set
as $T=300$K. The time unit $\tau$ is accordingly altered following
the formula applied for an amino acid \cite{Thirumalai1996} and
other details of the simulation process are the same as for the
folding and binding (see the following subsections). Based on 100
runs of molecular dynamics simulations starting from randomly
positions of all the particles and monomers in the box, average time
for the two monomers to diffuse into the confined space with
different sizes $L$ is obtained. A periodic boundary condition is
used to model the whole cell.

\subsection{Topology Based Model of the Homodimer}
A G\={o}-like potential is used to model the interactions within the
Arc homodimer. For each monomeric chain the interactions include the
virtual bonds $V^{s}_{bond}$, angles $V^{s}_{bond-angle}$, dihedral
angles $V^{s}_{dihedral}$, and non-bonded pairs of the C$_{\alpha}$
atoms $V^{s}_{non-bond}$ [for details see Ref.\cite{Clementi2000})].
Here the superscript $s$ denotes chain-A or chain-B, respectively.
Note that similar non-bonded interactions are also used for the
native and nonnative contacts between the inter-chain residues. The
native contact is defined as occurring when the distance between any
a pair of non-hydrogen atoms belonging to two residues is shorter
than $5.5$\AA in the native conformation of the dimer. Thus, the
monomeric and interfacial native contacts can be defined. In
addition, the crowding effect introduces a repulsive potential
$V^{c}(r_i)$ between the residue $i$ and the cylindrical wall when
their distance $r_i$ is less than $\sigma_{0}=4$\AA. Here
$V^{c}(r_i) = \sum_i 50 [(\sigma_0/2r_{i})^{4}
-2(\sigma_0/2r_{i})^{2}+1] \Theta(\sigma_0/2-r_{i})$ (for details
see Ref.\cite{excluded1}).

\subsection{Simulations for folding and binding} The simulations
were carried out using Langevin dynamics and leap-frog
algorithm\cite{Thirumalai1996,Levy2008}.  The native Arc dimer
is unfolded and equilibrated at high temperature, and then the
unfolded conformations are taken as starting states for the folding
simulations. The energy scale $\epsilon =1$ and time step $\delta
t=0.005 \tau$ are used. Here $\tau =\sqrt {m a^2 /\epsilon}$ is the
time scale with the van der Waals radius of the residues $a=5$\AA.
All the length is scaled by $\lambda =3.8 \AA$, i.e., the bond
length between two C$_{\alpha}$ atoms. A friction coefficient
$\gamma_{a} \sim 0.05$ is used. The thermodynamic variables (e.g.,
the free energy $F(Q)=E(Q)-T\log W(Q)$ with $E(Q)$ and $W(Q)$ are
energy of the system and the density of conformations at $Q$,
respectively) are obtained using the weighted histogram analysis
method (WHAM) \cite{Clementi2000}. The free energies for a monomeric
chain $F(Q_A)$ (or $F(Q_B)$ and for the chain-chain binding
$F(Q_{AB})$ can be calculated.

\end{materials}

\begin{acknowledgments}
This work was supported by the National Basic Research Program of
China (2006CB910302, 2007CB814806), and also the NNSF (10834002).
The support of the Center for Theoretical Biological Physics (NSF
PHY-0822283) is also gratefully acknowledged.
\end{acknowledgments}

\end{article}



\end{document}